\documentclass[twocolumn, amsmath, amssymb, preprintnumbers, showpacs, showkeys, aps, prb]{revtex4-1}

\usepackage{graphicx}
\usepackage{float}
\usepackage{color}
\begin{document}
\preprint{}

\title{Lone Pair Effect, Structural Distortions and Potential for Superconductivity in Tl Perovskites}

\author{Leslie M. Schoop$^{1,2}$}
\email{lschoop@princeton.edu}
\author{Lukas M\"uchler$^3$}
\author{Claudia Felser$^{3,4}$}
\author{R. J. Cava$^1$}

\affiliation{$^1$Department of Chemistry, Princeton University, Princeton New Jersey 08544, USA.}
\affiliation{$^2$Graduate School Material Science in Mainz, 55099 Mainz, Germany.}
\affiliation {$^3$Max-Planck-Institut f\"ur Chemische Physik fester Stoffe, 01187 Dresden, Germany.}
\affiliation{$^4$Institut f\"ur Anorganische und Analytische Chemie,
              Johannes Gutenberg - Universit\"at, 55099 Mainz, Germany.}
\date{\today}

\begin{abstract}

Drawing the analogy to BaBiO$_3$, we investigate via ab-initio electronic structure calculations potential new superconductors of the type ATlX$_3$ with A = Rb, Cs and X = F, Cl, and Br, with a particular emphasis on RbTlCl$_3$. Based on chemical reasoning, supported by the calculations, we show that Tl-based perovskites have structural and charge instabilities driven by the lone pair effect, similar to the case of BaBiO$_3$, effectively becoming A$_2$Tl$^{1+}$Tl$^{3+}$X$_6$. We find that upon hole doping of RbTlCl$_3$, structures without Tl$^{1+}$, Tl$^{3+}$ charge disproportionation become more stable, although the ideal cubic perovskite, often viewed as the best host for superconductivity, should not be the most stable phase in the system. The known superconductor (Sr,K)BiO$_3$ and hole doped RbTlCl$_3$, predicted to be most stable in the same tetragonal structure, display highly analogous calculated electronic band structures.

\end{abstract}


\maketitle

%%%%%%%%%%%%%%%%%%%%%%%%%%%%%%%%%%%%%%%%%%%%%%%%%%%%%%%%%%%%%%%%%%%%%%
\section{Introduction}

Although not anticipated as such, relativistic effects play an important role in all parts of chemistry. Among the most prominent examples are the color of Au and the fact that Hg is a liquid at ambient conditions. \cite{pyykko2012relativistic} These observations are commonly attributed to the inert pair effect (IPE), which is the contraction and stabilization of the 6s orbitals in the context of the Lanthanide contraction and relativistic effects.
The relativistic stabilization of the 6s orbitals greatly influences the chemistry of the heavy \textit{p}-block elements. TlCl$_3$ (Tl$^{3+}$ 6s$^0$), for example, readily decomposes to form TlCl (Tl$^{1+}$ 6s$^2$) and Cl$_2$. Relativistic effects are not only important in inorganic chemistry but also in materials science and condensed matter physics. The unique physics of topological insulators is due to compounds with heavy atoms that show strong relativistic effects. \cite{muchler2012topological} For a review on relativistic effects in general see reference 1. Here we study the relationship between superconductivity and charge instabilities due to the IPE in perovskite related structures. We focus on Tl compounds in particular, since Tl is the element with the most pronounced IPE known, and as a consequence is most stable in its +1 (6s$^2$) rather than its +3 (6s$^0$) oxidation state.\\

After the cuprates, the next highest T$_c$ perovskite superconductor family is based on BaBiO$_3$, \cite{baumert1995barium} which has a formal charge of +4 for Bi (which would be 6s$^1$). From a chemical perspective, the correct formula for BaBiO$_3$ should be Ba$_2$Bi$_2$O$_6$ with both Bi$^{3+}$ and Bi$^{5+}$ present. This 'charge disproportionation' is due to the IPE stabilization of the 6s$^2$ configuration of Bi$^{3+}$. 
BaBiO$_3$ crystallizes in a monoclinic distorted version of the perovskite structure; \cite{baumert1995barium} the BiO$_6$-octahedra are tilted such that the angle between them is 159.9$^\circ$ in contrast to 180$^\circ$ in the perfect perovskite structure. Due to the Bi$^{5+}$ and Bi$^{3+}$ present, larger and smaller BiO$_6$ octahedra are found. This distortion can also be described as an alternating breathing in and breathing out of the BiO$_6$ octahedra.
The charge separation in BaBiO$_3$ is widely accepted in the literature. \cite{cox1979mixed,sleight1993high,takegahara1994electronic,meregalli1998electron,baumert1995barium}
If BaBiO$_3$ is hole doped, either with K on the Ba site, or Pb on the Bi site, it becomes superconducting. The maximum T$_c$ is 34~K for Ba$_{0.6}$K$_{0.4}$BiO$_3$. \cite{cava1988superconductivity} Ba$_{1-x}$K$_x$BiO$_3$ (BKBO) crystallizes in the nominally simple cubic perovskite structure in the superconducting composition range 0.375 $<$ x $<$ 0.5 without any evidence of a breathing mode distortion, suggesting that simple cubic symmetry is preferred for the superconductivity. \cite{PhysRevB.41.4126} A more recent report, however, claims tetragonal symmetry for BKBO, the distortion arising from rotations of the octahedra. \cite{PhysRevB.62.6708} This is consistent with the structure of the Pb doped version. BaPb$_{1-x}$Bi$_x$O$_3$ (BPBO) is superconducting for 0.05 $<$ x $<$ 0.3 and its maximum T$_c$ is 13~K for x = 0.25 where there is a mixture of orthorhombic and tetragonal polymorphs. \cite{climent2011polymorphism} In both systems, BKBO and BPBO, the maximum T$_c$ is close to a metal to semiconductor transition.
When BaBiO$_3$ is hole doped, more Bi$^{5+}$ is present and the average Bi-O distance consequently decreases. Calculations have shown that a decreasing Bi-O distance causes the band gap to decrease and leads to metallic conductivity when enough electrons are removed. \cite{baumert1995barium} This yields superconductivity at a critical concentration of Bi$^{5+}$. The idea that these types of valence instabilities could lead to superconductivity has long been argued, \cite{larsson2003mixed,felser1997valence} and recent models quantify the presence of enhanced electron-phonon coupling due to this effect. \cite{yin2011correlation} \\ 
 
Here we show that the Tl perovskite RbTlCl$_3$, and by inference the larger class of hypothetical Tl perovskites (Rb,Cs)Tl(F,Cl,Br)$_3$, display the same electronic features as BaBiO$_3$. The nominal Tl$^{2+}$ present is expected to charge disproportionate into Tl$^{3+}$ and Tl$^{1+}$ due to the IPE, making RbTlCl$_3$ a charge density wave (CDW) insulator like BaBiO$_3$ and implying that superconductivity should be possible with hole doping. The expected T$_c$s of two hypothetical Cs variants with the ideal cubic perovskite structure in fact have recently been predicted within the framework of the local density approximation (LDA), using linear response theory (LRT). \cite{yin2013rational} Here we show that a van Hove singularity and a Fermi surface virtually identical to those seen in BaBiO$_3$ are present in the ideal cubic variant of the cubic Tl halide perovskites, but we also show by total energy calculations that the ideal non-charge-disproportionated cubic structure should never be expected in these systems even when hole doped. Rather, lower symmetry perovskites with rotational distortions are favored. Such phases are nonetheless also good hosts for superconductivity, as for example has been observed for (Sr,K)BiO$_3$. \cite{bougerol2000structural}

\section{Computational details}
The calculations were performed in the framework of density functional theory (DFT) using the \textsc{wien2k} \cite{blaha2001} code with a full-potential
linearized augmented plane-wave and local orbitals [FP-LAPW + lo] basis
\cite{singh2006,madsen2001,sjaestedt_alternative_2000} together with the Perdew Burke Ernzerhof (PBE) parameterization \cite{perdew_generalized_1996} of the generalized gradient approximation (GGA) as the
exchange-correlation functional. The plane wave cut-off parameter
R$_{MT}$K$_{MAX}$ was set to 7 and the irreducible Brillouin zone was sampled by 40-100 k-points. For each structure type the lattice constants were estimated with the program SPuDS, \cite{Lufaso:br0106} and optimized by minimizing the total energy. The atomic positions were optimized by minimization of the forces.

\section{Results}

The thallium halide perovskite CsTlCl$_3$ is a known compound which was reported once in the literature; \cite{ackermann2001gemischtvalente} the other possible compounds are not so far reported.
To estimate whether the formation of a perovskite should be possible, as well as its degree of distortion, the Goldschmitt tolerance factor $t$  is a good first step, \cite{Goldschmidt} where  $r_X$, $r_A$ and $r_M$ = anion radius, A site ion radius, and B site ion radius, respectively:
\begin{equation}
t = \frac{r_A + r_X}{\sqrt 2 \lbrack r_M + r_X \rbrack}
\label{eq:1}
\end{equation}
For a cubic structure, geometry requires that $t$ = 1. However, the ideal cubic structure occurs for 0.89~$<~t~<$~1. Distorted perovskites, where the octahedra tilt and rotate around their shared corners to accommodate the size of the ion in the cavity, appear for 0.8~$<~t~<$~0.89. For smaller $t$'s or $t>$~1, other structure types are usually favorable. However, pervoskites are not truly ionic and the value of the tolerance factor depends on the value for the radii, which are not fixed for non ionic compounds. The tolerance factor can therefore only be used as a rough estimate.
Using the Shannon radii, \cite{shannon1976revised} we calculate the perovskite tolerance factor $t$ (equation 1) for various Tl-based halide perovskites. 
The radius of Tl was estimated to be the average of the radius of Tl$^{1+}$ and Tl$^{3+}$. The results are shown in Table 1. For comparison, the tolerance factor for BaBiO$_3$ is 0.94.

\begin{table}[H]
\begin{center}
\caption{Tolerance factors for (Rb,Cs)Tl(F,Cl,Br)$_3$.}
\begin{tabular}{c|c}
\textbf{Compound} & $t$ \\
\hline
\hline
\textbf{RbTlF$_3$}  & 0.86\\
\textbf{RbTlCl$_3$} & 0.83\\
\textbf{RbTlBr$_3$} & 0.82\\
\textbf{CsTlF$_3$}  & 0.90\\
\textbf{CsTlCl$_3$} & 0.87\\
\textbf{CsTlBr$_3$} & 0.86\\
\end{tabular}
\end{center}
\label{tab_2}
\end{table}

The tolerance factors indicate that all of the compounds can crystallize in the perovskite structure, however, except for CsTlF$_3$, they are all expected to be distorted through rotations of the octahedra. If the Tl compounds are hole doped on either the halide or the alkali metal site, the radius of Tl ($r_M$) will become smaller because more Tl$^{3+}$ would be present. According to equation 1 the tolerance factor increases with decreasing $r_M$. Hence higher symmetry structures should be possible with hole doping.\\

We performed calculations for a model Tl halide perovskite, RbTlCl$_3$, in four different structure types: (a) an ideal undistorted cubic version (space group Pm$\bar{3}$m, no. 221), (b) a version with tilting of the octahedra as the only distortion (space group I$\bar{4}$mcm, no. 140, the symmetry of superconducting (Sr,K)BiO$_3$ \cite{bougerol2000structural}) (c) a structure with breathing mode distortion without tilting (space groups Fm$\bar{3}$m (225), the usual cubic double perovskite symmetry) and (d) one with both the breathing mode distortion and tilting of the octahedra (R$\bar{3}$ (148), the symmetry of high temperature BaBiO$_3$ \cite{zhou2004high}). All structures are presented in Figure 1. 

%%%%%%%%%%%%%%%%%%%%%%%%%%%%%%%%%%%%%%%%%%%%%%%%%%%%%%%%%%%%%%%%%%%%%%
\begin{figure}[h]
  \centering
  \includegraphics[width=0.45\textwidth]{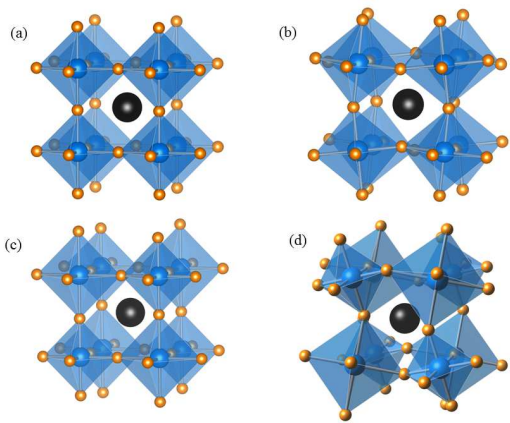}
  \caption{The four different structure types of RbTlCl$_3$ whose electronic structures are considered. a) the undistorted ideal cubic version, b) the tetragonal version with octahedra tilting as the only distortion (no breathing mode distortion), c) the cubic version with breathing modes and no tilting, d) the rhombohedral structure with both distortions, tilting and breathing modes. Big spheres represent Rb atoms, medium one Tl atoms and small ones Cl atoms.}
\label{fig_1}
\end{figure}
%%%%%%%%%%%%%%%%%%%%%%%%%%%%%%%%%%%%%%%%%%%%%%%%%%%%%%%%%%%%%%%%%%%%%%

Comparison of the total energies of the four possible RbTlCl$_3$ structure types shows that structures with breathing mode distortions, i.e. the presence of IPE-driven charge disproportionation, are much more stable than structures without breathing modes (Figure 2). As expected, we found such structures to be insulators with a band gap of the order of 1~eV. Structures without breathing modes calculate to be metallic without any doping, reinforcing the idea that the charge disproportionation stabilizes the structure by opening a band gap (Figure 3).

  %%%%%%%%%%%%%%%%%%%%%%%%%%%%%%%%%%%%%%%%%%%%%%%%%%%%%%%%%%%%%%%%%%%%%%
\begin{figure}[h]
  \centering
  \includegraphics[width=0.45\textwidth]{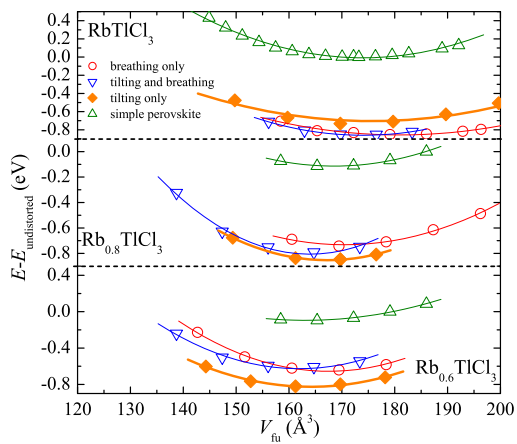}
  \caption{Total energies of RbTlCl$_3$ for different doping levels in the four structure types as  a function of cell volume.}
\label{fig_3}
\end{figure}
%%%%%%%%%%%%%%%%%%%%%%%%%%%%%%%%%%%%%%%%%%%%%%%%%%%%%%%%%%%%%%%%%%%%%%
 %%%%%%%%%%%%%%%%%%%%%%%%%%%%%%%%%%%%%%%%%%%%%%%%%%%%%%%%%%%%%%%%%%%%%%
\begin{figure}[h]
  \centering
  \includegraphics[width=0.45\textwidth]{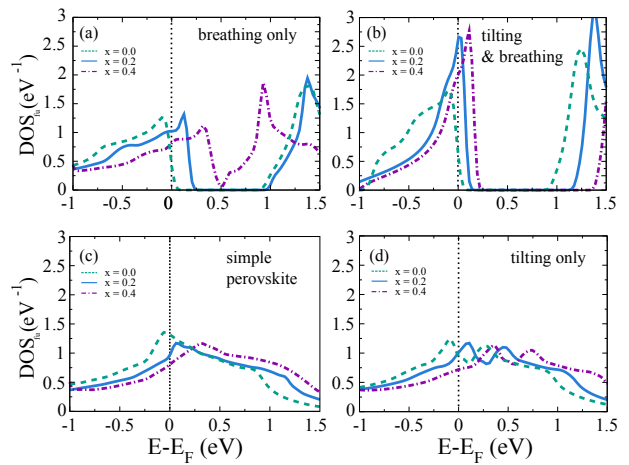}
  \caption{DOS of RbTlCl$_3$ for different doping levels in the four structure types.}
\label{fig_4}
\end{figure}
%%%%%%%%%%%%%%%%%%%%%%%%%%%%%%%%%%%%%%%%%%%%%%%%%%%%%%%%%%%%%%%%%%%%%%

 The band structure of the hypothetical undistorted ideal cubic structure of RbTlCl$_3$ shows a distinct van Hove singularity (vHS) at the X point (see Figure 4a), which is highly analogous to the one seen in BaBiO$_3$. \cite{takegahara1994electronic} The presence of the vHS at E$_F$ is indicative of a structural instability, where a distortion to decrease the density of states at E$_F$ is strongly favored. The vHS has been cited as important for the occurrence of superconductivity in perovskites and other materials. \cite{felser1997valence} It also suggests that a structural distortion is necessary to stabilize this compound, driven by the need to decrease the density of states (DOS) at E$_F$ that is present due to the vHS in the ideal cubic phase. The very strong electronic analogy of undoped ideal cubic RbTlCl$_3$ to undoped ideal cubic BaBiO$_3$ can be seen by comparison of the Fermi surfaces in figures 4b and 4c.    
 
 %%%%%%%%%%%%%%%%%%%%%%%%%%%%%%%%%%%%%%%%%%%%%%%%%%%%%%%%%%%%%%%%%%%%%%
\begin{figure}[h]
  \centering
  \includegraphics[width=0.45\textwidth]{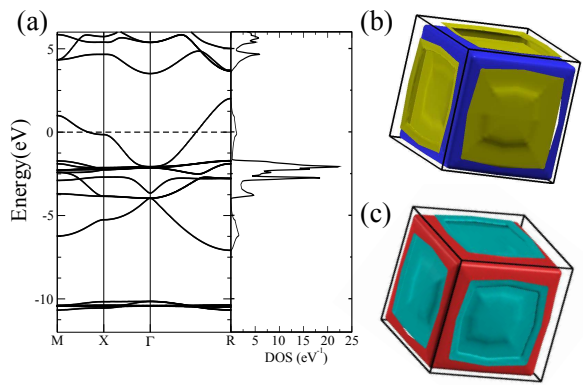}
  \caption{(a) Band structure and density of states of RbTlCl$_3$ in the cubic undistorted structure. A distinct vHS is visible at the X point. (b) Fermi surfaces of RbTlCl$_3$ and (c) BaBiO$_3$.}
\label{fig_2}
\end{figure}
%%%%%%%%%%%%%%%%%%%%%%%%%%%%%%%%%%%%%%%%%%%%%%%%%%%%%%%%%%%%%%%%%%%%%%
 
In order to simulate the effects of hole doping in RbTlCl$_3$, we performed calculations within the virtual crystal approximation, with fewer electrons on the Rb site, allowing us to simulate the hypothetical hole-doped compounds Rb$_{0.8}$TlCl$_3$ and Rb$_{0.6}$TlCl$_3$. We find that the volume decreases in each structure type in accordance with the increase of the Tl$^{3+}$ concentration. Hole doping closes the band gaps in the semiconducting structures. Comparison of the total energies shows that with increasing hole concentrations the energy differences between the structures with and without breathing modes becomes smaller. Figure 2 shows plots of energy versus volume for the different structures, for doped and undoped compounds. For the undoped compound, the breathing mode structures have a significantly lower energy than structures with regular octahedra, even for the case of dramatically decreased cell volumes. Therefore an ideal cubic perovskite structure should not be stable for undoped RbTlCl$_3$ for experimentally accessible pressures, a reflection of the very strong IPE. With a hole doping level of 0.2 per formula unit, the tetragonal structure with no breathing modes becomes lower in energy, and, at hole dopings of 0.4 per formula unit, the tilted, non charge disproportionated structure is clearly the most stable. Therefore, with hole doping the breathing mode distortion becomes unfavorable, and a higher symmetry is preferred, although again the ideal cubic phase should not be seen. A similar scenario is observed in BKBO, where, at a doping level of 0.4 holes per formula unit, there is no evidence for breathing mode distortions, and a recent paper reveals a tilted-only structural distortion for the superconducting crystal structure. \cite{PhysRevB.62.6708} In addition, we performed calculations for CsTlF$_3$, the hypothetical perovskite in the thallium halide family with the highest tolerance factor ($t$ = 0.9) to test whether it might adopt the ideal cubic structure with doping or pressure. However the energy curves were qualitatively similar to those seen for RbTlCl$_3$, from which we conclude that the ideal cubic structure should not be stable for this family as a whole.

In Figure 5 the DOS in the vicinity of E$_F$ for the four structures with and without doping is shown. The two structures with breathing mode distortions become metallic with doping, but the DOS at the Fermi level is very high. In the tetragonal structure, the DOS is much lower for Rb$_{0.6}$TlCl$_3$, which again suggests the disappearance of breathing modes with doping. On doping, the energy gained by Tl disproportionation becomes smaller and the tilting of the octahedra becomes more important. This indicates that disproportionation is more favorable if the average oxidation number is +2, i.e. that equal amounts of Tl$^{1+}$ and Tl$^{3+}$ are present and the material is insulating. Hole doping with x = 0.4 changes the average oxidation number to 2.4, which makes the stabilization by charge disproportionation less effective and should lead  to identical bond lengths in the different TlCl$_6$ octahedra. The increase of the unfavorable 6s$^0$ high oxidation state of Tl$^{3+}$ could then lead to electronic instabilities that eventually induce superconductivity.

To further draw the comparison between RbTlCl$_3$ and the bismuthates, we calculated the band structure of Sr$_{0.4}$K$_{0.6}$BiO$_3$, the composition with the highest T$_c$ in the (Sr,K)BiO$_3$ system, within the virtual crystal approximation at the appropriate electron count Sr$_{0.7}$BiO$_3$. Sr$_{0.4}$K$_{0.6}$BiO$_3$ crystallizes in the same tetragonal crystal structure that we chose to model the tilting only structure for RbTlCl$_3$.
Figure 5 shows the comparison of the band structures of tetragonal Rb$_{0.6}$TlCl$_3$ and Sr$_{0.4}$K$_{0.6}$BiO$_3$. The band structures are very similar, reinforcing the strong electronic analogy between Bi-O and Tl-Halide perovskites.

 %%%%%%%%%%%%%%%%%%%%%%%%%%%%%%%%%%%%%%%%%%%%%%%%%%%%%%%%%%%%%%%%%%%%%%
\begin{figure}[h]
  \centering
  \includegraphics[width=0.45\textwidth]{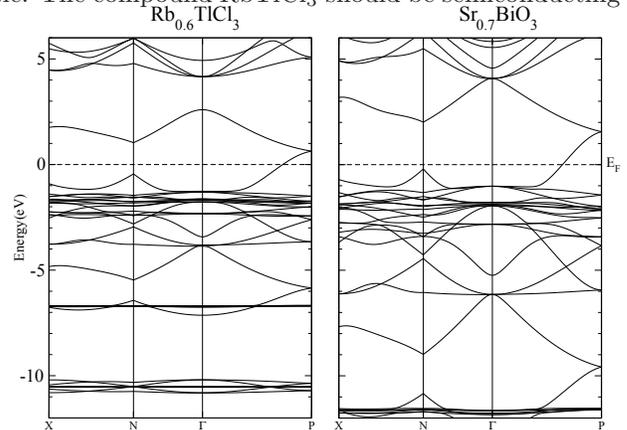}
  \caption{Comparison between the band structures of Rb$_{0.6}$TlCl$_3$ (left) and Sr$_{0.7}$BiO$_3$ (right).}
\label{fig_5}
\end{figure}
%%%%%%%%%%%%%%%%%%%%%%%%%%%%%%%%%%%%%%%%%%%%%%%%%%%%%%%%%%%%%%%%%%%%%%

%%%%%%%%%%%%%%%%%%%%%%%%%%%%%%%%%%%%%%%%%%%%%%%%%%%%%%%%%%%%%%%%%%%%%%
\section{Summary and conclusions}

Because the predicted most stable undoped form of RbTlCl$_3$ has frozen-in breathing modes, the scenario of charge disproportionation in Tl due to the IPE is realistic. The compound RbTlCl$_3$ should be semiconducting if undoped, just like BaBiO$_3$. Hole doping makes the band structure metallic and results in a stabilization of higher symmetry structures without frozen breathing modes, hence the possibility of superconductivity is present, in analogy to BaBiO$_3$. The electron-phonon coupling, an additional important factor for the occurrence of superconductivity, has also been predicted to be favorable in thallium halide perovskites \cite{yin2013rational}, complementing this picture.  

Other compounds in the family ATlX$_3$ with A = Rb, Cs, and X = F, Cl, Br should have similar features. Following the idea that the inert pair effect is the major reason for superconductivity in BKBO, one should also look for superconductivity in compounds in which Pb has an average oxidation number of +3. This could be present for example in CsPbCl$_2$O or BaPbO$_2$F.

%%%%%%%%%%%%%%%%%%%%%%%%%%%%%%%%%%%%%%%%%%%%%%%%%%%%%%%%%%%%%%%%%%%%%%
\bigskip 
\begin{acknowledgments}

The work at Princeton was supported by the Department of Energy grant DE-FG02-98-ER45706 and AFOSR grant FA9550-09-1-0593. RJC acknowledges work with L.F. Mattheiss, J.J. Kratewski and W.F. Peck, Jr. on this family of compounds in the early 1990s

\end{acknowledgments}

%%%%%%%%%%%%%%%%%%%%%%%%%%%%%%%%%%%%%%%%%%%%%%%%%%%%%%%%%%%%%%%%%%%%%%
\bibliography{Lit}

\end{document}